\documentclass[fleqn,twoside]{article}%
\usepackage{espcrc2}
\usepackage{graphicx,color}
\usepackage{amsmath,amssymb}
\usepackage[figuresright]{rotating}
\usepackage{amsmath}
\usepackage{amsfonts}
\usepackage{amssymb}
\usepackage{graphicx}%

\begin{document}
\pagestyle{empty}
\title{Calculating multiloop integrals using dimensional
recurrence relation and $\mathcal{D}$-analyticity}
\author{R.N. Lee \address[BINP]{The Budker Institute of Nuclear Physics}
\thanks{Talk given at "Loops and Legs in Quantum Field Theory" 25-30 April 2010, W\"orlitz, Germany}
}

\begin{abstract}
We review the method of the calculation of multiloop integrals recently
suggested in Ref.\cite{Lee2010}. A simple method of derivation of the
dimensional recurrence relation suitable for automatization is given. Some new
analytic results are given.

\end{abstract}
\maketitle

\section{Introduction}

Recently, in Ref \cite{Lee2010} a method of multiloop integrals evaluation
based on $\mathcal{D}$ recurrence relations \cite{Tarasov1996} and
$\mathcal{D}$-analyticity was suggested (DRA method). In this contribution we
give a brief review of this method. We also provide a simple method of
derivation of the dimensional recurrence relation well-suited for automatization.

\section{DRA method}

The DRA method has been described in detail in Ref. \cite{Lee2010}. It
consists of the following steps:

\begin{enumerate}
\item Make sure all master integrals in subtopologies are known. If it is not
so, start from calculating them.

\item Pass to a suitable master integral $J^{(\mathcal{D})}$. It is convenient
to choose a master integral which is finite in the basic stripe. For this
purpose, e.g., increase powers of some massive propagators.

\item Construct the dimensional recurrence relation for this master integral. The
general form of this recurrence is
\begin{equation}
J^{(\mathcal{D}-2)}=C(\mathcal{D})J^{(\mathcal{D})}+R(\mathcal{D}),
\end{equation}
where $C(\mathcal{D})$ is some rational function and $R(\mathcal{D})$ is a
non-homogeneous part constructed of the master integrals of subtopologies in
$\mathcal{D}$ dimensions.

\item Find a general solution of this recurrence relation
\begin{equation}
J^{(\mathcal{D})}=\Sigma^{-1}(\mathcal{D})\omega(z)+J_{\text{ih}%
}^{(\mathcal{D})},
\end{equation}
where $J_{\text{ih}}^{(\mathcal{D})}$ is a specific solution of the
inhomogeneous equation, $\Sigma^{-1}(\mathcal{D})$ is the solution of the
homogeneous equation, and $\omega(z)=\omega(\exp(i\pi\mathcal{D}))$ is
arbitrary periodic function.

\item Fix the singularities of $\omega(z)$ by analysing the analytical
properties of the master integrals and summing factor $\Sigma(\mathcal{D})$.

\item If needed, fix the remaining constants from the value of the integral at
some space-time dimension $\mathcal{D}$.
\end{enumerate}

Step 5 is the key point of the DRA approach. In order to perform this step one
needs to determine the position and order of the poles of $\Sigma
(\mathcal{D})J^{(\mathcal{D})}$ on suitable vertical stripe of width $2$ in
the complex plane of $\mathcal{D}$ (\textit{basic stripe}). This information
can be extracted from the parametric representation of the integral, either
manually or (semi-)automatically using the \texttt{FIESTA} code \cite{SmiSmTe2009}%
. In general, the number and the order of poles essentially depend on the
choice of the master integral $J^{(\mathcal{D})}$, the summing factor
$\Sigma(\mathcal{D})$, and the basic stripe. The proper choice may essentially
simplify the last step of the approach.

Before we proceed to the example of the application of the DRA method, we would
like to derive formulas convenient for the automatic derivation of the
dimensional recurrence relation.

\section{Dimensional recurrence relation}

The original derivation of the dimensional recurrence relation \cite{Tarasov1996}
is based on the parametric representation. For the integral without numerator
which can be represented by some graph the final formula has the form of some
sum over the graph trees. For the automatic calculation it may be desirable to
have the possibility to obtain the dimensional recurrence relation without any
reference to the graph and/or to the parametric representation. In this
Section we obtain the corresponding formulas using the Baikov's approach which
consists of the "changing of integration variables" from loop momenta to
scalar products (or denominators) \cite{Baikov1997}. We briefly review the
derivation of the corresponding transformation keeping also $\mathcal{D}%
$-dependent factors omitted in the original derivation of Ref.
\cite{Baikov1997}.

Assume that we are interested in the calculation of the $L$-loop integral
depending on $E$ linearly independent external momenta $p_{1},\ldots,p_{E}$.
There are $N=L(L+1)/2+LE$ scalar products depending on the loop momenta
$l_{i}$:%

\begin{equation}
s_{ij}=s_{ji}=l_{i}\cdot q_{j}\,;\quad i=1,\ldots,L;\quad j=1,\ldots,K,
\end{equation}
where $q_{1,\ldots,L}=l_{1,\ldots,L}$, $q_{L+1,\ldots,K}=p_{1,\ldots,E}$, and
$K=L+E$.

The loop integral has the form%
\begin{align}
J^{\left(  \mathcal{D}\right)  }\left(  \mathbf{n}\right)  =  &  \int
\frac{d^{\mathcal{D}}l_{L}\ldots d^{\mathcal{D}}l_{1}}{\pi^{L\mathcal{D}/2}%
}j(n_{1},\ldots,n_{N})\nonumber\\
=  &  \int\frac{d^{\mathcal{D}}l_{L}\ldots d^{\mathcal{D}}l_{1}}%
{\pi^{L\mathcal{D}/2}D_{1}^{n_{1}}D_{2}^{n_{2}}\ldots D_{N}^{n_{N}}}
\label{eq:J}%
\end{align}
where the scalar functions $D_{\alpha}$ are linear polynomials with respect to
$s_{ij}$. The functions $D_{\alpha}$ are assumed to be linearly independent
and to form a complete basis in the sense that any non-zero linear combination
of them depends on the loop momenta, and any $s_{ik}$ can be expressed in
terms of $D_{\alpha}$.

The integral $J^{\left(  \mathcal{D}\right)  }\left(  \mathbf{n}\right)  $ can
be considered as a function of $N$ integer variables. It is convenient
\cite{Lee2010} to introduce the operators $A_{i},B_{i}$ which act on such
functions as
\begin{align}
(A_{i}f)(\ldots,n_{i},\ldots)  &  =n_{i}f(\ldots,n_{i}+1,\ldots)\nonumber\\
(B_{i}f)(\ldots,n_{i},\ldots)  &  =f(\ldots,n_{i}-1,\ldots)
\end{align}

Let us first transform the innermost integral $\int d^{\mathcal{D}}l_{1}%
/\pi^{\mathcal{D}/2}$ in Eq. (\ref{eq:J}). The integrand $j$ depends on
$l_{1}$ via the scalar products $s_{1i}$ ($i=1\ldots L+E$). Writing
$l_{1}=l_{1\parallel}+l_{1\perp}$, where $l_{1\parallel}$ is the projection of
$l_{1}$ on the hyperplane spanned by $q_{2}\ldots q_{K}$, we obtain
\begin{align}
\frac{d^{\mathcal{D}}l_{1}}{\pi^{\mathcal{D}/2}} &  =\frac{d^{\mathcal{D}%
-K+1}l_{1\perp}}{\pi^{(\mathcal{D}-K+1)/2}}\frac{d^{K-1}l_{1\parallel}}%
{\pi^{(K-1)/2}}\nonumber\\
&  =\frac{\left(  \mu\frac{V(q_{1},\ldots q_{K})}{V(q_{2},\ldots q_{K}%
)}\right)  ^{(\mathcal{D}-K-1)/2}ds_{11}}{\Gamma\lbrack(\mathcal{D}%
-K+1)/2]}\nonumber\\
&  \times\frac{ds_{12}\ldots ds_{1K}}{\pi^{(K-1)/2}\sqrt{\mu^{K-1}%
V(q_{2},\ldots q_{K})}}\,
\end{align}
where $V(q_{1},\ldots q_{K})=\det\{s_{ij}|_{i,j=1\ldots K}\}$ is a Gram
determinant constructed on the vectors $q_{1},\ldots q_{K}$ and $\mu=\pm1$ for
the Euclidean/pseudoEuclidean case, respectively. Note that
\[
V\left(  q_{1},\ldots,q_{K}\right)  =P\left(  D_{1},\ldots,D_{N}\right)
\]
is a $K$-degree polynomial of $D_{\alpha}$.

Repeating the same transformation for $l_{2},\ldots,l_{L}$, we finally obtain
\begin{multline}
J\left(  \mathbf{n}\right)  =\frac{\pi^{\left(  L-N\right)  /2} \mu
^{L\mathcal{D}/2-N}}{\Gamma\left[  \left(  \mathcal{D}-K+1\right)
/2,\ldots,\left(  \mathcal{D}-E\right)  /2\right]  }\\
\hspace{-10mm}\times\int\left(  \prod_{i=1}^{L}\prod_{j=i}^{K}ds_{ij}\right)
\frac{[V\left(  q_{1},\ldots,q_{K}\right)  ]^{(\mathcal{D}-K-1)/2} }
{[V\left(  p_{1},\ldots,p_{E}\right)  ]^{(\mathcal{D}-E-1)/2} }j\left(
\mathbf{n}\right)  \label{eq:Baikov}%
\end{multline}

In order to use this formula in explicit calculations, we also need to
determine the limits of integration over the $s_{ij}$ variables. However, for
algebraic manipulations we only need to keep in mind that the integration by
part does not generate any surface terms.

The lowering dimensional recurrence relation is immediately obtained by
replacing $\mathcal{D}\rightarrow\mathcal{D}+2$ in Eq. (\ref{eq:Baikov}) and
comparing the resulting expression with the original one \cite{Lee2010}. We obtain%

\begin{multline}
J^{\left(  \mathcal{D}+2\right)  }\left(  \mathbf{n}\right)  =\frac{(2\mu
)^{L}\left[  V\left(  p_{1},\ldots,p_{E}\right)  \right]  ^{-1}}{\left(
\mathcal{D}-E-L+1\right)  _{L}}\\
\times\left(  P\left(  B_{1},\ldots,B_{N}\right)  J^{\left(  \mathcal{D}%
\right)  }\right)  \left(  \mathbf{n}\right)  . \label{eq:Lowering}%
\end{multline}

In order to obtain the relation between master integrals, we have to use IBP
reduction for the right-hand side of Eq. (\ref{eq:Lowering}). The complexity
of this reduction strongly depends on the integrals appearing in the
right-hand side. The lowering dimensional recurrence relation
(\ref{eq:Lowering}) contains integrals with indices shifted by at most $K=L+E$
in comparison with the integral in the left-hand aside.

The raising dimensional recurrence relation is more "economic" from this point
of view. In the original Tarasov's derivation the parametric representation of
the loop integral was used. For the integral given by some graph, the result
is expressed in terms of the trees of this graph. However, for the automatic
derivation of the raising recurrence relation this formula may be
inconvenient. Therefore, it is desirable to be able to obtain the raising
recurrence relation without any reference to the graph. In order to obtain the
raising recurrence relation we use the identity
\begin{multline}
\det\left\{  2^{\delta_{ij}}\frac{\partial}{\partial s_{ij}}|_{i,j=1\ldots
L}\right\}  [V\left(  q_{1},\ldots,q_{K}\right)  ]^{\alpha}\\
=(2\alpha)_{L}V\left(  p_{1},\ldots,p_{E}\right)  [V\left(  q_{1},\ldots
,q_{K}\right)  ]^{\alpha-1}\label{eq:diff}%
\end{multline}
The proof of this identity is based on the Carl Jacobi theorem about determinants and will be
presented elsewhere. The raising dimensional recurrence relation is obtained by replacing
$\mathcal{D}\rightarrow\mathcal{D}-2$ in Eq. (\ref{eq:Baikov}), substituting $V\left(
p_{1},\ldots,p_{E}\right)  [V\left(  q_{1},\ldots ,q_{K}\right)  ]^{(\mathcal{D}-K-3)/2}$ with the
derivative and integrating by part. We obtain
\begin{multline}
\hspace{-10mm}J^{\left(  \mathcal{D}-2\right)  }\left(  \mathbf{n}\right)
=(-\mu/2)^{L}\\
\hspace{-10mm}\times\int\frac{d^{\mathcal{D}}l_{L}\ldots d^{\mathcal{D}}l_{1}%
}{\pi^{L\mathcal{D}/2}}\det\left\{  \frac{2^{\delta_{ij}}\partial}{\partial
s_{ij}}|_{i,j=1\ldots L}\right\}  j\left(  \mathbf{n}\right)  \\
\hspace{-10mm}=(\mu/2)^{L}\\
\hspace{-10mm}\times\left(  \det\left\{  2^{\delta_{ij}}\frac{\partial D_{k}%
}{\partial s_{ij}}A_{k}|_{i,j=1\ldots L}\right\}  J^{\left(  \mathcal{D}%
\right)  }\right)  \left(  \mathbf{n}\right)  .\label{eq:Raising}%
\end{multline}

Comparing Eq. (\ref{eq:Raising}) with Tarasov's formula we obtain for the case
of integral corresponding to some graph:%
\[
\det\left\{  2^{\delta_{ij}-1}\frac{\partial D_{k}}{\partial s_{ij}}%
A_{k}|_{i,j=1\ldots L}\right\}  =\sum_{\text{trees}}A_{i_{1}}\ldots A_{i_{L}},
\]
where the sum goes over all trees of the graph, and $i_{1},\ldots,i_{L}$
enumerate the chords of the tree.

\section{Example}

Let us demonstrate the application of the method on the calculation of the
following four-loop vacuum integral:%

\begin{multline*}
\hspace{-10mm}J^{(\mathcal{D})}=
 \raisebox{-0.4525cm}{\includegraphics[ trim=0.000000in 0.000000in
-0.004115in 0.001443in, height=1.0632cm, width=1.0741cm
]%
{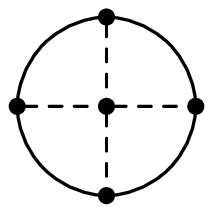}%
}%
\\
\hspace{-10mm}=\int\frac{d^{\mathcal{D}}k\,d^{\mathcal{D}}l\,d^{\mathcal{D}}r\,d^{\mathcal{D}}%
p}{\pi^{2\mathcal{D}}k^{2}l^{2}r^{2}\left(  k+l+r\right)  ^{2}\left[  \left(
p-k-l\right)  ^{2}+1\right]  }\\
\hspace{-10mm}\times\frac{1}{[\left(  p-k\right)  ^{2}+1][p^{2}+1][\left(  p+r\right)
^{2}+1]}%
\end{multline*}

This integral has been considered in Refs. \cite{SchrVuo2005,KiriLee2009}. In
Ref. \cite{SchrVuo2005} this integral has been evaluated numerically using the
Laporta's defference equation method. In Ref. \cite{KiriLee2009} this integral
has been considered using the dimensional recurrence relation. However, in
that paper in order to fix the periodic function parametrizing the homogeneous
solution we had to resort to the Laporta's difference equation. Here we
present the derivation entirely based on the DRA method. This derivation
serves solely as the illustration of the DRA method. The final result for
arbitrary $\mathcal{D}$ coincides with the result of Ref. \cite{KiriLee2009}.

\begin{enumerate}
\item There are four master integrals in the subtopologies:
\begin{align}
\hspace{-10mm}J_{1}^{\left(  \mathcal{D}\right)  }  &  \equiv
\raisebox{-0.4525cm}{\includegraphics[
trim=0.000000in 0.000000in -0.003066in -0.003060in,
height=1.0017cm,
width=1.0741cm
]%
{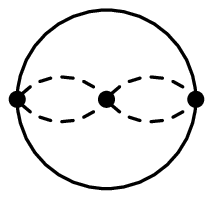}%
}%
,\quad J_{2}^{\left(  \mathcal{D}\right)  }=%
\raisebox{-0.4525cm}{\includegraphics[
trim=0.000000in 0.000000in -0.003066in -0.003060in,
height=1.0017cm,
width=1.0741cm
]%
{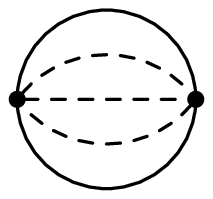}%
}%
,\nonumber\\
\hspace{-10mm}J_{3}^{\left(  \mathcal{D}\right)  }  &  =%
\raisebox{-0.7029cm}{\includegraphics[
trim=0.000000in 0.000000in -0.004386in 0.003552in,
height=1.5706cm,
width=1.0741cm
]%
{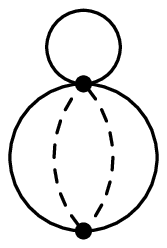}%
}%
,\quad J_{4}^{\left(  \mathcal{D}\right)  }=%
\raisebox{-0.4525cm}{\includegraphics[
trim=0.000000in 0.000000in -0.002337in -0.002337in,
height=1.0741cm,
width=1.0741cm
]%
{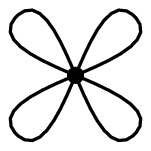}%
}%
. \label{eq:MIsimple}%
\end{align}
These integrals are expressed in terms of $\Gamma$-functions, see, e.g., Ref.
\cite{KiriLee2009}.

\item The integral has no ultraviolet divergence for $\mathcal{D}<4$. At
$\mathcal{D}=4$ the integral has a simple pole. The integral has a simple pole
also at $\mathcal{D}=2\frac{2}{3}$ due to the infrared infrared divergence.
This divergence comes from the region where $k,l,$ and $r$ are small.
Subtracting from the integrand the quantity $\left[  k^{2}l^{2}r^{2}\left(
k+l+r\right)  ^{2}\left(  p^{2}+1\right)  ^{4}\right]  ^{-1}$, corresponding
to a scaleless integral, we easily establish, that $J^{(\mathcal{D})}$ is finite
when $2<$Re$\mathcal{D}<2\frac{2}{3}$. We choose the basic stripe as
$S=\left\{  \mathcal{D}|\quad\operatorname{Re}\mathcal{D\in}\left(
2,4\right]  \right\}  $.

\item The dimensional recurrence for $J^{\left(  \mathcal{D}\right)  }$ reads%
\begin{multline}
\hspace{-10mm}J^{\left(  \mathcal{D}+2\right)  }=-\frac{48(3\mathcal{D}%
-11)(3\mathcal{D}-7)}{(\mathcal{D}-3)_{4}(\mathcal{D}-2)^{2}}J^{\left(
\mathcal{D}\right)  }\nonumber\\
\hspace{-10mm}+c_{1}^{\left(  \mathcal{D}\right)  }J_{1}^{\left(
\mathcal{D}\right)  }+c_{2}^{\left(  \mathcal{D}\right)  }J_{2}^{\left(
\mathcal{D}\right)  }+c_{3}^{\left(  \mathcal{D}\right)  }J_{3}^{\left(
\mathcal{D}\right)  }+c_{4}^{\left(  \mathcal{D}\right)  }J_{4}^{\left(
\mathcal{D}\right)  },
\end{multline}
where $c_{i}^{\left(  \mathcal{D}\right)  }$ are some rational functions not
presented here for brevity (see Ref. \cite{KiriLee2009}).

\item The summing factor obeys the equation
\begin{equation}
\frac{\Sigma\left(  \mathcal{D}\right)  }{\Sigma\left(  \mathcal{D}+2\right)
}=-\frac{48(3\mathcal{D}-11)(3\mathcal{D}-7)}{(\mathcal{D}-3)_{4}%
(\mathcal{D}-2)^{2}}%
\end{equation}
Since $J^{\left(  \mathcal{D}\right)  }$ has simple poles in the basic stripe
at $\mathcal{D}=2\frac{2}{3},4$, we choose the summing factor to have zeros at
these points. Namely, we choose%
\begin{equation}
\Sigma\left(  \mathcal{D}\right)  =\frac{\cos\left(  \frac{\pi\mathcal{D}}%
{2}+\frac{\pi}{6}\right)  \Gamma^{2}(\mathcal{D}-3)\Gamma\left(
\frac{\mathcal{D}}{2}\right)  }{8^{\mathcal{D}}\Gamma(3-\mathcal{D}%
)\Gamma\left(  \frac{3\mathcal{D}}{2}-\frac{11}{2}\right)  }\label{eq:I3}%
\end{equation}

The general solution of the dimensional recurrence has the form%
\begin{align}
\Sigma\left(  \mathcal{D}\right)  J^{\left(  \mathcal{D}\right)  } &
=\omega\left(  z\right)  -\sum_{i=1}^{4}s_{i}\left(  \mathcal{D}\right)  \\
s_{i}\left(  \mathcal{D}\right)   &  =\sum_{k=0}^{\infty}t_{i}\left(
\mathcal{D}+2k\right)  \label{eq:general}\\
t_{i}\left(  \mathcal{D}\right)   &  =\Sigma\left(  \mathcal{D}\right)
c_{i}^{\left(  \mathcal{D}\right)  }J_{i}^{\left(  \mathcal{D}\right)  }%
\end{align}

\item The left-hand side of Eq. (\ref{eq:general}) has no singularities on
$S$, so the right-hand side should also be a holomorphic function. The
functions $t_{i}\left(  \mathcal{D}+2k\right)  $ have poles at $\mathcal{D}%
=\mathcal{D}_{1-6}$ where
\begin{align*}
\mathcal{D}_{1} &  =2\frac{1}{2},\mathcal{D}_{2}=2\frac{2}{3},\mathcal{D}%
_{3}=3,\\
\mathcal{D}_{4} &  =3\frac{1}{3},\mathcal{D}_{5}=3\frac{1}{2},\mathcal{D}%
_{6}=4,
\end{align*}
The pole structure of $t_{i}\left(  \mathcal{D}+2k\right)  $ is demonstrated
in Fig. 1. Note that $s_{i}\left(  \mathcal{D}\right)  $ may have poles only
in the points where the individual terms $t_{i}\left(  \mathcal{D}+2k\right)
$ are singular.%
\begin{figure}
[ptb]
\begin{center}
\includegraphics[
width=3in
]%
{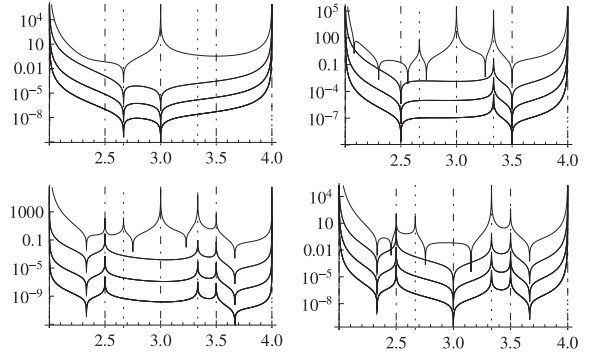}%
\caption{Pole structure of functions $t_{1},t_{2}$ (upper row), $t_{3},t_{4}$
(lower row). We plot $\left\vert t_{i}\left(  \mathcal{D}+2k\right)
\right\vert $ for $k=0,\ldots,3$.}%
\end{center}
\end{figure}

Therefore, we obtain%
\begin{equation}
\omega\left(  z\right)  =\sum_{l=1}^{6}\sum_{r=1}^{r_{l}}c_{l}^{r}\,\left[
\cot\left(  \frac{\pi}{2}\left(  \mathcal{D}-\mathcal{D}_{l}\right)  \right)
\right]  ^{r}+\mathrm{const},\label{eq:om}%
\end{equation}
where $r_{l}$ is the order of the pole at $\mathcal{D}=\mathcal{D}_{l}$, and
the coefficients $c_{l}^{r}$ should be chosen so as to cancel all
singularities in the right-hand side of Eq. (\ref{eq:general}). Their
determination is reduced to the solution of some linear system which we do not
present here for brevity.

As it is shown in Ref. \cite{Lee2010}, an $L$-loop integral is bounded in the
limit $\mathcal{D}\rightarrow\pm i\infty$ by $z^{\pm L/4}\left\vert \log
z\right\vert ^{\nu}$, where $\nu$ is some irrelevant exponent. Using this fact
and the explicit form of $\Sigma\left(  \mathcal{D}\right)  $, it is easy to
establish that $\Sigma\left(  \mathcal{D}\right)  J^{\left(  \mathcal{D}%
\right)  }$ and $s_{i}\left(  \mathcal{D}\right)  $ fall down when
$\mathcal{D}\rightarrow\pm i\infty$. Therefore, the constant in Eq.
(\ref{eq:om}) should be chosen in such a way that $\omega\left(  z\right)  $
falls down when $z\rightarrow0,\infty$. The first term in Eq. (\ref{eq:om})
has different limits when $\mathcal{D}\rightarrow\pm i\infty$, and we obtain%
\begin{equation}
\mathrm{const}=-\sum_{l=1}^{6}\sum_{r=1}^{r_{l}}c_{l}^{r}\left(  -i\right)
^{r}=-\sum_{l=1}^{6}\sum_{r=1}^{r_{l}}c_{l}^{r}\left(  +i\right)
^{r}\label{eq:const}%
\end{equation}
Eqs. (\ref{eq:general}),(\ref{eq:om}), and (\ref{eq:const}) entirely determine
$J^{\left(  \mathcal{D}\right)  }$ for arbitrary $\mathcal{D}$. However, we
may want to find the coefficients $c_{l}^{r}\ $in Eq. (\ref{eq:om})
explicitely. Using the fast convergence of sums in $s_{i}\left(
\mathcal{D}\right)  $ and keeping in mind the possibility to use the \texttt{pslq}
algorithm \cite{FergBai1991}, we find that values of all these coefficients
are compatible with zero, at least, up to $10^{-500}$. Therefore, we conclude
that
\begin{equation}
\omega\left(  z\right)  \overset{500}{=}0,\label{eq:om0}%
\end{equation}
where $\overset{500}{=}$ denotes the equality checked numerically with $500$ digits.

\item Our consideration allowed us to fix all constants within the method.
Adopting the guess (\ref{eq:om0}), we obtain%
\[
J^{\left(  \mathcal{D}\right)  }=-\Sigma^{-1}\left(  \mathcal{D}\right)
\sum_{i=1}^{4}s_{i}\left(  \mathcal{D}\right)
\]

\end{enumerate}

This result coincides with that of Ref. \cite{KiriLee2009}. Using the
\texttt{pslq} algorithm, we can express the expansion near $\mathcal{D}=4$ in
terms of conventional $\zeta$-values:%

\begin{align*}
&  \hspace{-1cm}J^{\left(  4-2\epsilon\right)  }\overset{300}{=}\frac
{\epsilon^{3}\Gamma\left[  -1+\epsilon\right]  ^{4}}{1+\epsilon}\left[
1+2\epsilon^{3}+3\epsilon^{4}+O\left(  \epsilon^{5}\right)  \right] \\
&  \hspace{-1cm}\times\left[  5\zeta_{5}-\left(  7\zeta_{3}^{2}+\frac
{11\pi^{6}}{378}\right)  \epsilon\right.  +\left(  \frac{\pi^{4}\zeta_{3}}%
{30}+212\zeta_{7}\right)  \epsilon^{2}\\
&  \hspace{-1cm}-\left(  \frac{29213\pi^{8}}{32400}-1820\zeta_{2,6}%
-5038\zeta_{3}\zeta_{5}\right)  \epsilon^{3}\\
&  \hspace{-1cm}+\left(  \frac{13255\zeta_{9}}{3}+\frac{731\pi^{4}\zeta_{5}%
}{6}-\frac{2006\pi^{6}\zeta_{3}}{189}+\frac{1006\zeta_{3}^{3}}{3}\right)
\epsilon^{4}\\
&  \hspace{-1cm}\left.  +O\left(  \epsilon^{5}\right)  \right]
\end{align*}

Note that the factor in the first line is chosen so as to provide the uniform
transcendentality weight in the rest of expansion. In Ref. \cite{SchrVuo2005}
the first term of the above expansion has been found analytically and the remaining
terms have been found numerically with 40-digit precision.

Let us also present the result for the expansion around $\mathcal{D}=3$, which
can be important for the calculations in hot QCD:%
\begin{align*}
&  \hspace{-1cm}J^{\left(  3-2\epsilon\right)  }\overset{300}{=}\frac
{\Gamma\left[  -1/2+\epsilon\right]  ^{4}}{1+\epsilon}\left[  \frac{\pi^{2}%
}{96}+\frac{11\zeta_{3}}{16}\epsilon\right.  \\
&  \hspace{-1cm}\left.  +\left(  \frac{271\pi^{4}}{2880}+\pi^{2}\log
2-\frac{5\zeta_{3}}{2}-\frac{41\pi^{2}}{48}\right)  \epsilon^{2}+O\left(
\epsilon^{3}\right)  \right]
\end{align*}
In Ref. \cite{KiriLee2009} the first term of the above expansion has been
found analytically and the second term has been obtained numerically.

\section{Conclusion}

We have briefly reviewed the method of calculation of multiloop integrals
based on the $\mathcal{D}$-recurrence and $\mathcal{D}$-analyticity. The
method appeares to be powerful enough to deal with the most complicated cases.
We have also derived convenient formulas, Eqs. (\ref{eq:Lowering}) and
(\ref{eq:Raising}), suitable for the automatic derivation of the Tarasov's
dimensional recurrence. For a specific four-loop master integral we have
presented in analytic form several terms of the expansion around $\mathcal{D}=4$
and $\mathcal{D}=3$.

This work was supported by RFBR (grants Nos. 07-02-00953, 08-02-01451) and DFG (grant
No. GZ436RUS113/769/0-2). I appreciate the organizers' support  for the participation in the
workshop. I also thank for warm hospitality the Max-Planck Institute for Quantum Optics,
Garching, where a part of this work was done.

\bibliographystyle{h-elsevier2}

\end{document}